\documentclass[
  ,draft            
  ]
  {aipproc}
\usepackage[cp850]{inputenc}
\usepackage{amsmath,amsfonts,amssymb,amsthm}
\usepackage[english]{babel}
\usepackage{latexsym}
\usepackage{amscd}
\usepackage{float}
\usepackage[T1]{fontenc}

\newcommand\ephin{\mathcal E(\Phi,\mathcal N)}
\newcommand\Pa{\mathcal P_a}

\newcommand\fnu{f_{\nu}}

\newcommand\lzero{\lambda_0}
\newcommand\cra{\curvearrowright}

\newcommand\la{\langle}
\newcommand\ra{\rangle}

\newcommand\ovlk{\overline{\kappa}}
\newcommand\tildeh{\tilde{\Hh}}
\newcommand\moneplus{M^1_+(\mathbb A)}
\newcommand{\plk}{\ln_{\alpha,\beta}}

\newcommand{\suf}{_{\alpha,\beta}}

\newtheorem{theorem}{Theorem}
\DeclareMathOperator{\val}{val}
\DeclareMathOperator{\HH}{H}
\DeclareMathOperator{\Hh}{h}
\DeclareMathOperator{\MaxEnt}{MaxEnt}
\DeclareMathOperator{\D}{D}
\DeclareMathOperator{\Dd}{d}

\layoutstyle{6x9}

\begin{document}

\title{Exponential Families and MaxEnt Calculations for Entropy
  Measures of Statistical Physics}

\classification{65.40.Gr}
\keywords      {Complexity, Game Theoretical Equilibrium, 
Maximum Entropy, Robustness, Exponential Families, Bregman Generator.}

\author{Flemming Tops\o e}{
  address={University of Copenhagen, Institute of
Mathematical Sciences\\
Universitetsparken 5, 2100 Copenhagen, Denmark}
}

\begin{abstract}
  For a wide range of entropy measures, easy calculation of equilibria
  is possible using a principle of {\it Game Theoretical Equilibrium}
  related to Jaynes {\it Maximum Entropy Principle}.  This follows
  previous work of the author and relates to Naudts \cite{naudts04a},
  \cite{naudts04b}, and, partly, Abe and Bagci \cite{abebagci05}.
 \end{abstract}

\maketitle


\section{The principle of Game 
Theoretical Equilibrium}\label{sect:1}

Consider a discrete {\it alphabet} $\mathbb A$ and {\it probability
  distributions}  $P, Q,\cdots$ over
$\mathbb A$. The set of all such distributions is denoted
$M^1_+(\mathbb A)$. A distribution is identified by its
point probabilities: $P=(p_i)_{i\in\mathbb A}$.
A  {\it measure of complexity} is a map
which to each pair $(P,Q)$ of distributions assigns a
value $\Phi(P,Q)\in[0,\infty]$ such that, for each $P\in
M^1_+(\mathbb A)$, the minimal value of $\Phi(P,Q)$ with $Q\in
M^1_+(\mathbb A)$ is assumed on the diagonal, i.e. for $Q=P$ and
nowhere else unless $\Phi(P,P)=\infty$.

A {\it preparation} is any non-empty subset $\mathcal P\subseteq
M^1_+(\mathbb A)$. When $\mathcal P$ is fixed, a {\it consistent
  distribution} is a distribution in $\mathcal P$. The {\it game}
$\gamma=\gamma(\Phi,\mathcal P)$ has $\Phi$ as objective function and
is the two-person zero-sum game between {\it Player I} (\lq\lq
Nature\rq\rq\,), who can choose a strategy $P\in\mathcal P$, and {\it
  Player II} (\lq\lq the Physicist\rq\rq\,) who can choose any
strategy $Q\in M^1_+(\mathbb A)$. Player I is a maximizer, Player II a
minimizer. Thus $\val_I$ defined by
$
\val_I=\sup_{P\in\mathcal P}\inf_{Q}\Phi(P,Q)
$
is the {\it Player I-value} of the game and, similarly, $\val_{II}$
defined by
$
\val_{II}=\inf_Q\sup_{P\in\mathcal P}\Phi(P,Q)
$
is the {\it Player II-value} of the game. Here and below,
a variable denoted by $Q$ is understood to vary over
all of $\moneplus$.

An {\it optimal Player I-strategy} is a $P\in\mathcal P$ such that
$\val_I=\inf_{Q}\Phi(P,Q)$ and an {\it optimal Player II-strategy} is
a $Q\in M^1_+(\mathbb A)$ such that $\val_{II}=\sup_{P\in\mathcal
  P}\Phi(P,Q)$. By the general {\it minimax inequality}, $\val_I\leq
\val_{II}$. The game is in {\it equilibrium} if
$\val_I=\val_{II}<\infty$.  

For further information about the game introduced,
see \cite{topsoenext}. 
The attempt to locate optimal strategies for the
players and to establish equilibrium for suitable preparations is
taken as a basic principle of statistical physics,
the {\it principle of game theoretical equilibrium}
(GTE).  

We introduce {\it $\Phi$-entropy} of
$P$ as minimal complexity, i.e. as
$\HH(P)=\inf_{Q}\Phi(P,Q)$. By assumption,
$
\HH(P)=\Phi(P,P)
$, thus, 
$
\val_I=\sup_{P\in\mathcal P}\HH(P)
$,
which is the {\it maximum entropy value}, also denoted
$\MaxEnt=\MaxEnt(\Phi,\mathcal P)$.  So $\val_I=\MaxEnt$ and we
realize that the GTE-principle leads directly to
Jaynes {\it maximum entropy
  principly}, cf. \cite{Jaynes57}.

Classical Boltzmann-Gibbs-Shannon entropy (BGS-entropy) is obtained as
minimal complexity with respect to the measure $(P,Q)\cra\sum
p_i\ln\frac{1}{q_i}$ which has a clear and convincing interpretation
related to coding.  Our results go some way to establish reasonable
interpretations also for more general measures of complexity.
Regarding the origin of the the above measure of complexity, under the
name of {\it inaccuracy}, see Kerridge \cite{Kerridge61}.


As we have seen, entropy is generated by complexity. So is {\it
  divergence  (cross entropy, relative entropy} or {\it
  redundancy)}, defined as actual minus minimal
complexity:
$
\D(P,Q)=\Phi(P,Q)-\HH(P)
$
when $\HH(P)<\infty$.  In any case, the {\it
  linking identity}
$
\Phi(P,Q)=\HH(P)+\D(P,Q)
$
holds and $\D(P,Q)\geq 0$ with equality if and only if $P=Q$
 (for the measures of
  complexity we shall consider, it will be clear how to define
  $\D(P,Q)$ when $\HH(P)=\infty$).

\section{Robustness, exponential 
families}\label{sect:2}

A Player II-strategy $Q$ is {\it robust} if, for some constant
$h<\infty$, the {\it level of robustness},
$\Phi(P,Q)=h$ for all consistent distributions $P$.  The set $\mathcal
E=\mathcal E(\Phi,\mathcal P)$ of all robust Player II-strategies is
the {\it exponential family} associated with $\gamma(\Phi,\mathcal
P)$. If a family $\mathcal N$ of preparations is considered, the
{\it exponential family} $\mathcal E(\Phi,\mathcal N)$ associated with
$\mathcal N$ is the set of distributions which are robust for all
preparations $\mathcal P\in\mathcal N$.

The following general and simple
observation will play a key role in the sequal:

\begin{theorem}[robustness lemma]\label{th:1}
  Let the measure of complexity $\Phi$ and the preparation $\mathcal
  P$ be given.  Assume that the distribution $Q^*$ is robust
  ($Q^*\in\mathcal E(\Phi,\mathcal P)$) and consistent
  ($Q^*\in\mathcal P$). Then
  $\gamma(\Phi,\mathcal P)$ is in equilibrium and has $Q^*$ as the
  unique MaxEnt-distribution as well as the unique optimal strategy
  for Player II.
\end{theorem}

\begin{proof}
  Though known from e.g. \cite{topsoenext} we present a direct proof.

  Let $h$ be the level of robustness. Then $\Phi(Q^*,Q^*)=h$ and, for
  $P\in\mathcal P$ with $P\neq Q^*$, $\HH(P)=\Phi(P,P)<\Phi(P,Q^*)=h$.
  Thus $Q^*$ is the unique MaxEnt-distribution. For any $Q\neq Q^*$,
  $\sup_{P\in\mathcal
    P}\Phi(P,Q)\geq\Phi(Q^*,Q)>\Phi(Q^*,Q^*)=h=\sup_{P\in\mathcal
    P}\Phi(P,Q^*)$ and equilibrium as well as unique optimality of
  $Q^*$ for Player II follows.
\end{proof}

The result connects the exponential family
$\mathcal E$ with the preparation $\mathcal P$. Indeed, if $\mathcal E$
and $\mathcal P$ intersect, they only intersect in one distribution
which then is the optimal strategy for both players
and, furthermore, the game considered is in equilibrium.

\section{Complexity and linear constraints}\label{sect:A}

We shall apply the principle of GTE -- via the
robustness lemma -- to a wide class of complexity functions and
associated notions of entropy, always having one and the same type
of preparations in mind, viz. those given by {\it linear constraints}.
They are the most important preparations for statistical physics and
other applications, cf. e.g.  Kapur \cite{kapur}.  

>From now on, we consider a fixed finite set
$f=(\fnu)_{1\leq\nu\leq k}$ of real-valued functions defined on
$\mathbb A$. The associated {\it family of natural preparations},
denoted $\mathcal N$, consists of all non-empty sets $\Pa$ which are
defined as follows, denoting by $\la\cdot,P\ra$ mean value w.r.t.
$P$:
\begin{equation}\label{eq:1}
 \Pa=\{P\in\moneplus\vert\la\fnu,P\ra=a_{\nu}\mbox{ for }1\leq\nu\leq k\}\,.
\end{equation}
Here $a=(a_{\nu})_{1\leq\nu\leq k}\in\mathbb R^k$.  We assume that no
non-trivial linear combination of the $\fnu$'s reduces to a constant
function.
Clearly, $\ephin$, the {\it natural exponential family},
consists of those distributions which are robust
for all natural preparations.

We shall select special measures of complexity adapted to a
study of the natural preparations and constructed with the aim
to simplify the search for distributions in
$\ephin$. To accomplish this, we consider measures of complexity of
the form
\begin{equation}\label{eq:3}
\Phi(P,Q)=\xi_Q\Big(\la\ovlk(Q),P\ra\Big)
\end{equation}
where, for each $Q\in M^1_+(\mathbb A)$, $\xi_Q$ is a real function
and $\ovlk$ maps $Q\in\moneplus$ into a function defined on $\mathbb
A$.  We insist that $\la\ovlk(Q),P\ra$ can be obtained by summation
based on a function $\kappa:[0,1]\to[0,\infty]$, the {\it coding
  function}, via the formula
\begin{equation}\label{eq:21}
\la\ovlk(Q),P\ra=\sum_{i\in\mathbb A}p_i\kappa(q_i)\,.
\end{equation}
This corresponds to the requirement
$(\ovlk(Q))(i)=\kappa(q_i)\,;\,i\in\mathbb A$.  

Regarding $\xi_Q:[0,\infty]\to[0,\infty]$ and
$\kappa:[0,1]\to[0,\infty]$, we assume that the $\xi_Q$'s are
increasing and concave, that $\kappa$ is decreasing and convex, that
$\kappa(1)=0$, that $\kappa$ is continuous at $0$ (not just at
$]0,1]$) and, finally, that $\Phi$ defined by \eqref{eq:3}
is a genuine measure of complexity. The last requirement will be trivially
fulfilled in the concrete cases we shall consider. The inverse
function $\kappa^{-1}:[0,\kappa(0)]\to[0,1]$ will play a significant
role. We note that this function is continuous, decreasing and convex,
as is $\kappa$ (simple geometric proof).

For the {\it classical example}, $\xi_Q$ is the identity map and
$\kappa$ the function $q\cra\ln\frac1q$. Then
$\kappa^{-1}$ is the restriction of $x\cra\exp{(-x)}$ to $[0,\infty]$.
Entropy generated by this measure of complexity is standard
BGS-entropy.

For the general situation, we note that
any $Q$ for which $\ovlk(Q)$ is a linear
combination of the constant function $1$ and the given functions
$f_1,\cdots,f_k$, i.e. of the form
\begin{equation}\label{eq:10}
\ovlk(Q)=\lzero+\lambda_1\cdot f_1+\cdots\lambda_k\cdot
f_k=\lzero+\lambda\cdot f
\end{equation}
for certain constants $\lzero$ and
$\lambda=(\lambda_1,\cdots,\lambda_k)$, is a member of $\ephin$. 
Motivated by this observation, we fix real constants
$\lambda=(\lambda_1,\cdots,\lambda_k)$ and ask if
there exists a real constant $\lambda_0$ and a distribution
$Q=(q_i)_{i\in\mathbb A}$ such that \eqref{eq:10} holds. 

For abbreviation, put $L_i=\lambda\cdot f(i)$. Then \eqref{eq:10}
amounts to $q_i=\kappa^{-1}(\lambda_0+L_i)$ for $i\in\mathbb A$. As
$\kappa^{-1}$ is defined on $[0,\kappa(0)]$, we must have
$0\leq\lambda_0+L_i\leq\kappa(0)$ for each $i$.  Therefore, the $L_i$
must be bounded below. Furthermore, from $\sum_iq_i=1$, we conclude
that, for each $K<\kappa(0)$, there can only be finitely many
$i\in\mathbb A$ with $L_i\leq K$. Thus we may order the $L_i$:
$L_{i_1}\leq L_{i_2}\leq\cdots$, with this sequence breaking off and
having a largest element if $\mathbb A$ is finite and with
$L_{i_n}\to\kappa(0)$ if $\mathbb A$ is infinite. Put $L_*=L_{i_1}$
and $L^*=\sup_{i\in\mathbb A}L_i$ ($=\kappa(0)$ if $\mathbb A$ is
infinite). We realize that we must require that $L^*-L_*\leq\kappa(0)$
and, assuming this holds, the set of possible constants $\lambda_0$ is
the set $[-L_*,\infty[$ in case $\kappa(0)=\infty$ and the set
$[-L_*,\kappa(0)-L^*]$ if $\kappa(0)<\infty$.  Consider the function
$f$ defined by
$
f(x)=\sum_{i\in\mathbb A}\kappa^{-1}(x+L_i)
$
with $x$'s ranging over the
possible values of $\lambda_0$. What we search for is a
value of $\lambda_0$, necessarily unique, such that $f(\lambda_0)=1$. 

Clearly, $f(-L_*)\geq 1$. By standard
techniques, we see that $f$ is continuous from
the right and if $f(x_0)<\infty$ for some value of $x_0$, then $f$ is
continuous at all $x>x_0$. Furthermore, if $x_n\to\kappa(0)$ and if
$f(x_n)<\infty$ for all $n$, then $f(x_n)\to 0$ as $n\to\infty$.

Our analysis shows that $f$ can have at most one point of
discontinuity, viz. where it passes from the value $\infty$ to finite
values. Such a discontinuity \lq\lq normally\rq\rq\, does not occur.
Also other anomalies are \lq\lq normally\rq\rq\, excluded. For
instance, one may easily construct examples such that $f$ is
constantly equal to $\infty$ but such values are also excluded as they
are of no practical interest. Thus we maintain that \lq\lq
normally\rq\rq\, the function $f$ assumes finite values larger than
$1$ as well as values less than $1$ and hence the existence of a value
$\lambda_0$ with $f(\lambda_0)=1$ is assured by continuity.

Summarizing, we can now
formulate the main result:

\begin{theorem}[MaxEnt calculus]\label{th:2}
  Let $\lambda=(\lambda_1,\cdots,\lambda_k)$ be given real constants.
  Then, under \lq\lq normal\rq\rq\, circumstances (cf. the discussion
  above),
the equation
\begin{equation}\label{eq:31}
\sum_{i\in\mathbb A}\kappa^{-1}\Big(\lambda_0+\lambda\cdot f(i)\Big)=1
\end{equation}
has a solution, necessarily unique, and $Q=(q_i)_{i\in\mathbb A}$ given by
\begin{equation}\label{eq:32}
q_i=\kappa^{-1}\Big(\lambda_0+\lambda\cdot f(i)\Big)\mbox{ for } i\in\mathbb
A
\end{equation}
satisfies \eqref{eq:10} and hence belongs to the exponential
family $\mathcal E(\Phi,\mathcal N)$.
This distribution is the MaxEnt-distribution for 
$\mathcal P_a$ with $a=(a_1,\cdots,a_k)$ given by
\begin{equation}\label{eq:33}
a_{\nu}=\sum_{i\in\mathbb A}q_if_{\nu}(i)\mbox{ for }\nu=1,\cdots,k
\end{equation}
 and, for this value of $a$,
$
\MaxEnt(\Phi,\mathcal P_a)=\xi_Q(\lambda_0+\lambda\cdot a)\,.
$
\end{theorem}

The theorem replaces and expands the standard recipe for
MaxEnt-calculations.  The main difference is a focus on $\lambda_0$
via \eqref{eq:31} rather than on the classical partition function. In
the final section we present a more thorough discussion of the
significance of the result.


Before continuing, we shall limit the type of complexity functions
studied by reducing the number of parameters needed for their
definition. Instead of the many functional parameters appearing in
\eqref{eq:3}, we now suggest
a setting with only two functional parameters, one function $\xi$,
called the {\it corrector}, to account for all the functions
$\xi_Q$ via the formula
$
\xi_Q(x)=x+\sum_{i\in\mathbb A}\xi(q_i)
$
and then the already introduced coding function $\kappa$.
In other words, we point to complexity functions of the
form 
\begin{equation}\label{eq:22}
\Phi(P,Q)=\sum_{i\in\mathbb A}p_i\kappa(q_i)+\sum_{i\in\mathbb A}\xi(q_i)\,.
\end{equation}

The functions $\kappa$ and $\xi$ are uniquely determined from $\Phi$.
The two terms in \eqref{eq:22} are called, respectively the {\it
  coding part} and the {\it correction}.  For the classical example,
the coding part is $\sum_ip_i\ln\frac1q$ and the correction vanishes.

\section{Complexity $\grave{{\bf A}}$ la Bregman}
\label{sect:3}

We shall now generate a $(\Phi,\HH,\D)$-triple from a simple starting
point. The method follows the idea of {\it Bregman divergences} and is
referred to as {\it Bregman generation}.  Another method, {\it
  Csisz{\'a}r generation},  was suggested in \cite{topsoenext}.  
In our view,
Bregman generation is by far the most important one for
the needs of statistical physics.

Given is a {\it Bregman generator} by which we shall understand a strictly
concave and smooth real function $\Hh$ defined on $[0,1]$ with
$\Hh(0)=\Hh(1)=0$ and 
$\Hh'(1)=-1$. We take \lq\lq smoothness\rq\rq\, to mean that $\Hh$ has
an analytic extension to $[0,\infty[$. Though less will do for most
investigations,
the stronger requirement allows one to consider also the {\it dual
  function} $\tildeh$ defined by
\begin{equation}\label{eq:80}
\tildeh(x)=x\Hh\Big(\frac1x\Big)\,.
\end{equation}
This function is well-defined and real-valued in $]0,\infty[$. As a
final technical assumption, we assume that the function can be
extended by continuity to $[0,\infty]$, allowing for infinite values
at the endpoints.
A specific value $\Hh(p)$ is interpreted as the {\it complexity} of an
event which is known to occur with probability $p$.

>From $\Hh$ we generate two functions,
$\phi=\phi(p,q)$, and  $\Dd=\Dd(p,q)$:
\begin{align}\label{eq:11}
\phi(p,q)&=\Hh(q)+(p-q)\Hh'(q)\,,\\\label{eq:20}
\Dd(p,q)&=\Hh(q)-\Hh(p)+(p-q)\Hh'(q)\,.
\end{align} 
A specific value $\phi(p,q)$ is interpreted as the {\it complexity} of
an event which is believed to occur with probability $q$ but actually
occurs with probability $p$. This is consistent with the previous
interpretation as $\phi(p,p)=\Hh(p)$. 
The function $\Dd$
simply measures the difference ({\it divergence}) between estimated
and true value.  We also note that $\phi(p,q)$ and $\Dd(p,q)$ 
may assume the value $+\infty$. This happens if and only if both $p>q=0$
and $\Hh'(0)=\infty$ hold.

Consider the {\it internal} functions, $\Phi=\Phi_{\Hh}$,
$\HH=\HH_{\Hh}$ and $\D=\D_{\Hh}$ generated by $\phi$, $\Hh$ and
$\Dd$. By this we mean that:
\begin{equation}\label{eq.12}
\Phi(P,Q)=\sum_{i\in\mathbb A}\phi(p_i,q_i)\,,\,\,
\HH(P)=\sum_{i\in\mathbb A}\Hh(p_i)\,,\,\,
\D(P,Q)=\sum_{i\in\mathbb A}\Dd(p_i,q_i)\,.
\end{equation}

We refer to $\phi$, $h$ and $d$ as the {\it partial} functions,
respectively partial {\it complexity, entropy} and {\it
  divergence}.
They satisfy a partial version of the linking identity:
\begin{equation}\label{eq:81}
\phi(p,q)=\Hh(p)+\Dd(p,q)\,.
\end{equation}

Note that $\Phi=\Phi_{\Hh}$ is of the special form \eqref{eq:22} with
coding function $\kappa=\kappa_{\Hh}$ given by
\begin{equation}\label{eq:24}
\kappa(x)=\Hh'(x)+1
\end{equation} 
and
corrector $\xi=\xi_{\Hh}$ given by 
$
\xi(x)=\Hh(x)-x(\Hh'(x)+1)$. Hence the Bregman generator is
decomposed into two terms:
\begin{equation}\label{eq:99}
\Hh(x)=x\kappa(x)+\xi(x)\,.
\end{equation} 

As $\xi(0)=\xi(1)=0$ and $\xi'(x)=-x\Hh''(x)-1$ we find
that $\xi\equiv 0$ if and only if we are in
the {\it classical case} $h(x)=x\ln (1/x)$. We also see that
$\xi(x)\geq-x$ in $[0,1]$, hence the correction related to any
distribution $Q$ is bounded below by $-1$.
The dual function $\tilde{\Hh}$ appears also to be of significance. In
particular, $\xi(x)=\tilde{\Hh}'(1/x)-x$,
hence 
\begin{equation}\label{eq:41}
\Phi(P,Q)=\sum_{i\in\mathbb A}p_i\Hh'(q_i)+\sum_{i\in\mathbb
  A}\tilde{\Hh}'(\frac{1}{q_i})\,.
\end{equation}
The first term in \eqref{eq:41} is the coding part minus $1$, the
second term the correction plus $1$.
Partial complexity is given by 
$
\phi(p,q)=p\Hh'(q)+\tilde{\Hh}'(1/q)
$.

\section{Generators via 
deformed logarithms}\label{sect:C}

We turn to a
concrete two-parameter family $(\Hh\suf)$ of Bregman generators
defined via {\it deformed logarithms} (taken
in this form from \cite{BorgesRoditi98}) and given by
\begin{equation}\label{eq:53}
\plk x=
\begin{cases}
\frac{x^{\beta}-x^{\alpha}}{\beta-\alpha}\mbox{ for }\alpha\neq\beta\\
x^{\alpha}\ln x\mbox{ for }\alpha=\beta
\end{cases}\,.
\end{equation}
The associated Bregman generators are defined by
\begin{equation}\label{eq:52}
\Hh\suf(x)=x\,\plk(1/x)\,.
\end{equation}
{\it Warning:} We have chosen to model the definition after the 
expression $x\,\ln(1/x)$
rather than $-x\,\ln x$. 
The main reason is
the more natural interpretation of the former expression, but also,
the change appears to be more as preferred in the \lq\lq Tsallis literature\rq\rq\,.
The change is in
contrast to the choice in \cite{topsoenext}. 
Thus, compared to \cite{topsoenext},
one should make the transformation $(\alpha,\beta)\cra(-\beta,-\alpha)$.
Note also the symmetry $\Hh\suf=\Hh_{\beta,\alpha}$.

>From \cite{topsoenext} we see (after transformation) that, in order to
obtain a genuine Bregman generator, the following restrictions apply
to $\alpha$ and $\beta$: {\it Either} $0\leq\alpha<1$ and $\beta\leq
0$ {\it or else} $\alpha\leq0$ and $0\leq\beta<1$.

The partial
complexity function and the coding function are given by:
\begin{align}\label{eq:82}
\phi\suf(x,y)&=\frac{1}{\beta-\alpha}
\Big(-(1-\alpha)x y^{-\alpha}+(1-\beta)x y^{-\beta}-
\alpha y^{1-\alpha}+\beta y^{1-\beta}\Big)\,,\\
\kappa\suf(x)&=1-\frac{1}{\beta-\alpha}\Big((1-\alpha)x^{-\alpha}-(1-\beta)x^{-\beta}\Big)\,.
\end{align}
Note that $\kappa(0)=\infty$ except if either $\alpha=0$ or $\beta=0$
(then $\kappa(0)=(\alpha+\beta-1)/(\alpha+\beta)$).

The important inverse functions $\kappa^{-1}$ are defined on
$[0,\kappa(0)]$. They can only be calculated in closed form in special
cases. We point to the {\it Tsallis case} which
corresponds to $\alpha<1$, $\beta=0$. The {\it Tsallis parameter},
traditionally denoted by $q$, is then given by $q=1-\alpha$. For the
origin to this family within the physics literature, see Tsallis,
\cite{Tsallis88}. Let us put $\kappa_{\alpha,0}=\kappa_q$ (as above
with $q=1-\alpha$). Then, for $q\neq 1$,
\begin{equation}\label{eq:60}
\kappa^{-1}_q(x)=\Big(1+\frac{1-q}{q}x\Big)^{\frac{1}{q-1}}\mbox{ for
}0\leq x\leq\kappa_q(0)
\end{equation}
and one can insert \eqref{eq:60} into \eqref{eq:31}. The kind of sums
obtained will, typically, have to be calculated numerically. An
exception is the case $q=2$. We leave it to the reader to work out the
pleasent details of our calculus in this case (take $\mathbb A$ to be finite).

Another case where $\kappa^{-1}\suf$ can be calculated in closed form
is the {\it Kaniadakis family} which corresponds to  
$\alpha=-\beta$, cf. Kaniadakis \cite{kaniadakis02}. We shall not go
into that here.

\section{Discussion}\label{sect:4}

{\it Some features of the main result.} Theorem \ref{th:2} provides a
theoretical framework for MaxEnt calculations for natural preparations
given by linear constraints and pertaining to a wide range of
different entropy measures. Among special features as compared with
the standard approach we mention the following:

The basis for the result is the game theoretical approach which
necessitates a focus on possibly unfamiliar aspects and quantities,
notably a focus on a notion of complexity, intended to reflect the
interplay between the physicist and the system he is studying.  This
aspect could have been hidden, but the underlying principle -- the
principle of {\it Game Theoretical Equilibrium} -- is in itself
promoted as a major issue. Indeed, it is suggested that this principle
is of a basic nature, applicable to several scientific investigations,
and that, for the area of statistical physics, it is more fundamental
than Jaynes Maximum Entropy Principle. The principle originated with
Pfaffelhuber \cite{Phuber77} and, independently, the author (with
\cite{Topsoe79} the first publication in English). Among further
studies, we mention the joint work \cite{Hartop01} with
Harremo{\"e}s.

Another feature is the puzzling fact that optimization has been
achieved \lq\lq miraculously\rq\rq\, without recourse to Lagrange
multipliers. Many will find it difficult to accept that for the
problem studied, an approach which is better -- simpler and more
illuminating -- than the well proven technique involving the popular
multipliers exists. Within the mathematical literature, this special
feature goes back at least to Csisz{\'a}r, cf.
\cite{Csiszar75}.

Finally, we note that the MaxEnt calculus outlined here has no mention
of partition functions.  The calculus goes a good deal
beyond traditional settings based on classical BGS-entropy. This has
resulted in a focus on $\lambda_0$ which corresponds to the logarithm
of the partition function in the classical case (so, for the classical
case, we can write $\lambda_0=\ln Z(\lambda)$ where $Z(\lambda)
=\sum\exp(-\lambda\cdot f(i))$). It is well known that $\ln Z$ is a key
quantity to work with, thus this feature should be no great surprise.
But it is interesting that our approach leads directly to this
quantity. As the partition function has no place for the general case
covered by Theorem \ref{th:2}, this is of course also forced in some
sense.

{\it Exponential families.}
Whereas the concept of partition function does not survive the
extension to general entropy- and complexity measures, the notion of
{\it exponential families} does. It even appears to be {\it the}
central concept behind the approach taken, cf. Theorem \ref{th:1}.
However, extensions of this concept are needed (see below).

{\it Comparing with the classical approach.}
The simplifications in the classical case result from the
factorization property of $\kappa^{-1}$, an exponential function in
that case. Apart from this, the
calculations for a general complexity function appear to be of much
the same nature as for the classical case. Indeed, given
$\lambda=(\lambda_1,\cdots,\lambda_k)$ one determines $\lambda_0$ from
\eqref{eq:31} and then, via \eqref{eq:32}, \eqref{eq:33} leads to the
relevant averages $a=(a_1,\cdots,a_k)$. If you aim for a specific set
of averages, there seems to be no way, neither in the classical case
nor in the general setting, other than application of numerical
optimization procedures to choose just that set of parameters
$\lambda$ which leads to the appropriate set of constrained values.
This discussion then tells us that apart from the simplifications
possible in handling \eqref{eq:31}, the general calculus suggested is
no more complicated in practise than what you are used to from
classical studies.

{\it Thermodynamic calculus.}
The difficulties, indeed impossibilities, involved in finding
solutions to MaxEnt problems in closed form for other than the
simplest problems constitute part of the motivation to create a
thermodynamic calculus, studying variation as functions of various
parameters of significance to the physicist or chemist. In this way
one hopes to develop useful approximate solutions or to discover
interesting trends in the thermodynamics as response to changes of
relevant parameters.
The differential calculus needed for such endeavours appears to be
applicable also to the general setting of Theorem \ref{th:2} with its
precise equations to look closer into. Studies of this kind are 
not taken up here.

{\it Natural expansions, optimal opdating based on a prior.}
There are many further possibilities for theoretical
investigations based on measures of complexity of the form here
studied. Assumptions related to the form
\eqref{eq:3} allows one to derive several results other than Theorem
\ref{th:2}: Uniqueness of $Q$ determined from $\lambda$, convexity of
the set of $\lambda$'s for which $Q$ can be found, convexity of the
function $\lambda\cra \lambda_0=\lambda_0(\lambda)$ (this corresponds
in the classical case to log-convexity of the partition function),
existence of equilibria for the models in the natural family and, as a
consequence, concavity of the map $a\cra$ MaxEnt$(\Phi,\mathcal P_a)$.

We comment that whereas measures of complexity of
the special form \eqref{eq:22} are rather simple and quite a rich family, the
more elaborate form given by \eqref{eq:3} is also of importance --
especially, it allows the consideration of {\it R{\'e}nyi entropies}
and related quantities.

A special expansion of the concept of robustness which allows
identification of MaxEnt-distributions for which some of the point
probabilities (the $q_i$ of Theorem \ref{th:2}) are allowed to be $0$
should also be mentioned. This concerns cases where
$\lambda_0+\lambda\cdot f(i)\geq\kappa(0)$ and is therefore only
relevant when $\kappa(0)<\infty$. However, there are important cases
where this is so, e.g. Tsallis-type quantities with $q>1$. In such
cases {\it inconsistent} inference is possible where a {\it feasible}
$i$ (one for which there exists $P\in\mathcal P_a$ with $p_i>0$) is
inferred under MaxEnt-based inference as an impossible event. This
phenomenon is treated in part by Jaynes, cf. p.345 of \cite{Jaynes03}.  Taking
this into consideration, it appears possible to prove
that any candidate to MaxEnt-distributions (or the more general {\it
  centers of attraction} of \cite{Hartop01}) of preparations in a
natural family of preparations, must be a member of the associated
exponential family. For the classical case, where inconsistent
inference is not possible, such a result was established in
\cite{Hartop01}.

Consider now the problem of optimal {\it updating} based on a
given {\it prior}. In fact, such problems can be handled in analogy
with our analysis of MaxEnt problems. In particular, a result
$\grave{{\rm a}}$ la Theorem \ref{th:2} holds which provides a
calculus for optimal {\it posterior distributions} via a {\it minimum
  cross entropy principle} --  the kind of results initiated by
Kullback, cf.  \cite{Kullback59}. To indicate, if only briefly, that
this requires no new techniques, consider a prior $Q_0$ and try to
maximize the {\it updating gain}
$\Psi_{Q_0}(P,Q)=\Phi(P,Q_0)-\Phi(P,Q)$. This situation can be
analyzed by applying our game theoretical reasoning to $-\Psi_{Q_0}$
which is a genuine complexity measure. For this to work, the theory
has to be extended slightly, allowing complexity measures that can
take negative values. 

Precise statements and proofs of results just indicated will
be published elsewhere.

{\it Origin of the two-parameter family.}
The two-parameter family of complexity-, entropy- and divergence
measures, $(\Phi\suf,\HH\suf,\D\suf)$ has its origin in the
mathematical literature, cf. Mittal \cite{Mittal75} and Sharma and
Taneja \cite{SharmaTaneja75}, and was studied later in the physical
literature by Borges and Roditi, \cite{BorgesRoditi98} who used the
convenient concept of deformed logarithms.

{\it Entropy should not stand alone.}
Let us illustrate this thesis
by considering Tsallis entropy with Tsallis parameter $q$. There are
infinitely many ways of obtaining this entropy measure as minimal
complexity. Below we suggest three complexity measures which have this
property:
\begin{align}\label{eq:70}
\Phi^B(P,Q)&=\frac{1}{q-1}+\sum\Big(q_i^q-\frac{q}{q-1}p_iq_i^{q-1}\Big)\\
\Phi^C(P,Q)&=\frac{1}{1-q}\sum p_i^q(1-q_i^{1-q})\\
\Phi^R(P,Q)&=\frac{1}{1-q}\Big(\frac{\sum p_i^q}{\sum
  p_i^qq_i^{1-q}}-1\Big)\,.
\end{align}

As usual, sums are over $i\in\mathbb A$.
The \lq\lq B \rq\rq\,, \lq\lq C\rq\rq\, and \lq\lq R\rq\rq\, stand
for, respectively \lq\lq Bregman \rq\rq\,, \lq\lq Csisz{\'a}r\rq\rq\,
and \lq\lq R{\'e}nyi\rq\rq\,. The complexity measure $\Phi^B$ is the one
considered in the main text, $\Phi^C$  the one considered in
\cite{topsoenext} and $\Phi^R$ is closely related to the
relevant complexity measure connected with R{\'e}nyi entropy and divergence.

The measure $\Phi^B$ allows us -- as we have seen -- to study the
natural preparations given by linear constraints, $\Phi^C$ allows us
to develop a calculus much as Theorem \ref{th:2}, but aiming at
maximizing entropy for preparations given by averaging 
with respect to the {\it
  $q$-associated measures} which are measures with point masses
$p_i^q$ and finally, $\Phi^R$ allows us to deal with preparations
given by averages with respect to the {\it $q$-escort distributions}
which are obtained by normalizing the $q$-associated measures. To
realize that this is indeed so, you just have to note how $P$ enters
in the complexity measure considered. It can safely be argued that
\lq\lq distorted\rq\rq\, averages as those indicated above related to
$\Phi^C$ and $\Phi^R$ have no physical relevance and therefore, they
are considered of less or no importance for the study of natural
maximum entropy problems. Bregman generation is thus the method which
stands back as the really significant method.

{\it The importance of Bregman type quantities.}
The relevance for statistical physics of Bregman divergence was
emphasized by Naudts \cite{naudts04a}, \cite{naudts04b}. The work by
Abe and Bagci \cite{abebagci05} should also be mentioned, however, the
present author does not agree with their conclusion that the use of
escort distributions is essential. Anyhow, the proper matching of
entropy measure with the type of constraints one wants to
study is important. This issue is also addressed in Feng \cite{Feng07}.

Originally, Bregman introduced the concept to meet needs of learning
theory, cf. \cite{Bregman67}. For more recent articles in this direction,
see Murata et al., \cite{Murataetal04} and Sears \cite{Sears07}.
 
Concerning extensions in another direction, to quantum statistical
physics, note the recent study by Petz,
\cite{Petz07} where Bregman divergences are carefully
defined. Incorporation of game theoretical considerations may be a
fruitful area of research to look into. 

{\it Interpretations.}
Any measure of entropy of importance to statistical physics should be
motivated by sound reasons, including appropriate interpretations. It
appears 
that Bregman generation in
itself goes a way in this direction. In addition, the choice of
terminology, especially regarding the frequent reference to \lq\lq
coding\rq\rq\,, though not yet founded in precise procedures for
observation or measurement, is indicative for what future research may
bring, at least this is where speculations of the author goes.

One should recall that Kullback-Leibler divergence is related to free
energy for classical preparations. This kind of interpretation when
more general Bregman-type divergences are involved appears also to be
sound, cf. the recent study by Bagci, \cite{Bagci07}.  Possibly,
Crooks, \cite{Crooks07}, also points to issues to be integrated before
a full picture is in place.




\begin{theacknowledgments}
Thanks are due to Jan Naudts for critical comments and suggestions and
to Bjarne Andresen for pointing me to \cite{Crooks07}.
The work was supported by the
    Danish Natural Science Research Council.
\end{theacknowledgments}

\end{document}